# A Hybrid Solution Method for the Capacitated Vehicle Routing Problem Using a Quantum Annealer

**Sebastian Feld**[*]
LMU Munich
Mobile and Distributed Systems Group
Munich, Germany

**Christoph Roch**
LMU Munich
Mobile and Distributed Systems Group
Munich, Germany

**Thomas Gabor**
LMU Munich
Mobile and Distributed Systems Group
Munich, Germany

**Christian Seidel**
Volkswagen Data:Lab
Munich, Germany

**Florian Neukart**
Volkswagen Group of America
San Francisco, USA

**Isabella Galter**
Volkswagen Data:Lab
Munich, Germany

**Wolfgang Mauerer**
OTH Regensburg
Regensburg, Germany

**Claudia Linnhoff-Popien**
LMU Munich
Mobile and Distributed Systems Group
Munich, Germany

## Abstract

The Capacitated Vehicle Routing Problem (CVRP) is an NP-optimization problem (NPO) that has been of great interest for decades for both, science and industry. The CVRP is a variant of the vehicle routing problem characterized by capacity constrained vehicles. The aim is to plan tours for vehicles to supply a given number of customers as efficiently as possible. The problem is the combinatorial explosion of possible solutions, which increases superexponentially with the number of customers. Classical solutions provide good approximations to the globally optimal solution. D-Wave's quantum annealer is a machine designed to solve optimization problems. This machine uses quantum effects to speed up computation time compared to classic computers. The problem on solving the CVRP on the quantum annealer is the particular formulation of the optimization problem. For this, it has to be mapped onto a quadratic unconstrained binary optimization (QUBO) problem. Complex optimization problems such as the CVRP can be translated to smaller subproblems and thus enable a sequential solution of the partitioned problem. This work presents a quantum-classic hybrid solution method for the CVRP. It clarifies whether the implementation of such a method pays off in comparison to existing classical solution methods regarding computation time and solution quality. Several approaches to solving the CVRP are elaborated, the arising problems are discussed, and the results are evaluated in terms of solution quality and computation time.

---

[*]sebastian.feld@ifi.lmu.de

# 1 Introduction

Optimization problems can be found in many different domains of applications, be it economics and finance [3], logistics [9], or healthcare [7]. Their high complexity engaged reseachers to develop efficient methods for solving these problems [35]. With D-Wave Systems releasing the first commercially available quantum annealer in 2011 [2], there is now the possibility to find solutions for such problems in a completely different way than classical computation does. To use D-Wave's quantum annealer the optimization problem has to be formulated as a quadratic unconstrained binary optimization (QUBO) problem [4], which is one of two input types acceptable by the machine (alternative: the Ising model [18]). Doing this, the metaheuristic quantum annealing seeks for the minimum of the optimization function, i.e., the best solution of the defined configuration space [31]. There has been recent research about solving real world problems on a quantum annealer, like Volkswagen's Traffic Flow Optimization [33] or the recently announced Tsunami Evacuation Optimization project by Tohoku University. [3]

The paper at hand regards the Capacitated Vehicle Routing Problem (CVRP), an NP-optimization problem that plays a major role in common operations research and is excessively studied since its proposal in [15]. The classic CVRP can be described as the problem of designing optimal routes from one depot to a number of geographically scattered customers subject to some side constraints (see Figure 1). It can be formulated as follows:

Let $G = (V, E)$ be a graph with $V = \{1, ..., n\}$ being a set of vertices representing $n$ customer locations with the depot located at vertex 1 and $E$ being a set of undirected edges. With every edge $(i, j) \in E, i \neq j$ a non-negative cost $c_{ij}$ is associated. This cost may, for instance, represent the (geographical) distance between two customers $i$ and $j$. Furthermore, assume there are $m$ vehicles stationed at the depot that have the same capacity $Q$. In addition, every customer has a certain demand $q$ [24]. The CVRP consists of finding a set of vehicle routes such that

- each customer in $V \setminus \{1\}$ is visited exactly once by exactly one vehicle;
- all routes start and end at the depot;
- the sum of customer demand within a route does not exceed the vehicles' capacity;
- the sum of costs of all routes is minimal given the constraints above;

To solve the CVRP on D-Wave's quantum annealer, the formulated QUBO problem has to be mapped to the hardware. However, quantum computation compared to classical computation is still in its infancy and one of the major problems is that quantum hardware is limited regarding the number of quantum bits (qubits) and their connectivity on the chip. Generally, this leads to difficulties in mapping large QUBO problems to the hardware.

With this paper we present an intuitive way to split the CVRP into smaller optimization problems by taking advantage of a classical 2-Phase-Heuristic [26], see Figure 1. This heuristic divides the CVRP into two phases, the clustering phase and the routing phase. The clustering phase itself can be mapped to the NP-complete Knapsack Problem (KP) [22], which tries to pack different sized items (here: customers) into capacity restricted knapsacks (here: vehicles). Doing this, the sum of the objective values of the items in a knapsack should be maximized, i.e. the euclidean distance between customers assigned to a vehicle should be minimized. The routing phase can be represented by the NP-hard Travelling Salesman Problem (TSP) [27]. Thus, the minimal tour in which all customers of a cluster are visited once is sought. Doing this, the tour starts and ends in one place, i.e. the depot. Figure 1 shows a CVRP example with the 2-Phase-Heuristic. First the customers are grouped into clusters (b) before efficient vehicle routes in each cluster are searched (c).

In this paper we investigate different quantum-classic hybrid approaches to solve the CVRP, expound their difficulties in finding good solutions, and finally propose a hybrid method based on the 2-Phase-Heuristic to solve the CVRP using D-Wave's quantum annealer. We map the optimization problems to a QUBO problem, and analyze performance from an application-specific perspective by using large benchmark datasets.

---

[2]https://www.dwavesys.com/news/d-wave-systems-sells-its-first-quantum-computing-system-lockheed-martin-corporation

[3]https://www.dwavesys.com/sites/default/files/Ohzeki.pdf

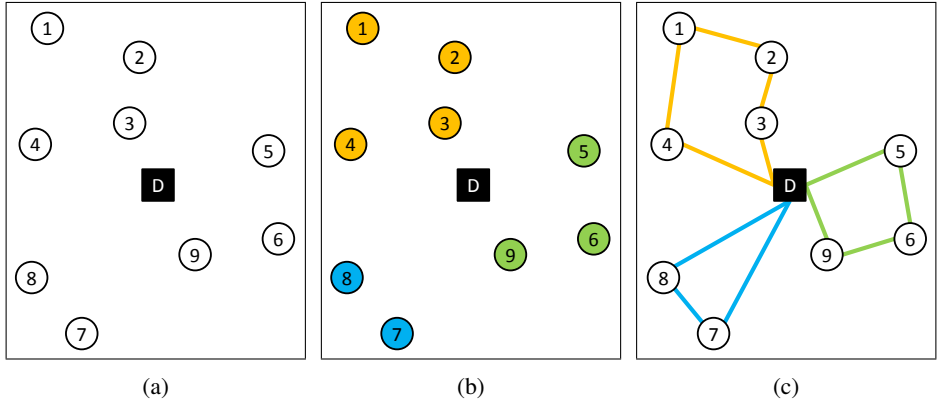

Figure 1: Overview of the CVRP and the 2-Phase-Heuristic. (a) Initial state with 9 customers and 1 depot. (b) Clustering phase results in three clusters found. (c) Routing phase determines shortest path inside each cluster.

The paper is structured as follows: Section 2 gives an introduction to quantum annealing and the common QUBO problem. A brief overview of existing methods for solving the CVRP is given in Section 3. In Section 4, two approaches for solving the CVRP are briefly discussed before the concept of our hybrid method is presented. Section 5 first introduces the test setup, and subsequently presents and discusses the results with regard to solution quality and computational performance on commonly used CVRP datasets. Finally, we conclude this paper in Section 6.

## 2 Quantum Annealing on D-Wave processor

Quantum annealing in general is a metaheuristic for solving complex optimization problems [21]. D-Wave's quantum annealing algorithm is implemented in hardware using a framework of analog control devices to manipulate a collection of quantum bit (qubit) states according to a time-dependent Hamiltonian, denoted $H(t)$, shown in Equation 1.

$$H(t) = s(t)H_I + (1 - s(t))H_P \tag{1}$$

The basic process of quantum annealing is to physically interpolate between an initial Hamiltonian $H_I$ with an easy to prepare minimal energy configuration (or ground state), and a problem Hamiltonian $H_P$, whose minimal energy configuration is sought that corresponds to the best solution of the defined problem. This transition is described by an adiabatic evolution path which is mathematically represented as function $s(t)$ and decreases from 1 to 0 [31]. If this transition is executed sufficiently slow, the propability to find the ground state of the problem Hamiltonian is close to 1 [1].

The just described concept of adiabatic quantum computing is the source of inspiration for the design of D-Wave's quantum annealing hardware. While the machine's functioning is based on following an adiabatic evolution path, the dynamics describing its working is not adiabatic. This is because the machine is strongly coupled to the environment resulting in the performance being affected by dissipative effects [30]. Nonetheless, the hardware is known to be capable of solving a specific optimization problem called a quadratic unconstrained binary optimization (QUBO) problem [4]. QUBO is a unifying model which can be used for representing a wide range of combinatorial optimization problems. However, in order to use quantum annealing on D-Wave's hardware the CVRP has to be formulated as a QUBO problem. The functional form of the QUBO the quantum annealer is designed to minimize is:

$$\min x^t Q x \qquad \text{with } x \in \{0, 1\}^n \tag{2}$$

with $x$ being a vector of binary variables of size $n$, and $Q$ being an $n \times n$ real-valued matrix describing the relationship between the variables. Given the matrix $Q$, the annealing process tries to find binary variable assignments to minimize the objective function in Equation 2.

The quantum processing unit (QPU) is a physical implementation of an undirected graph with qubits as vertices and couplers as edges between them. These qubits are arranged according in a so-called chimera graph, as illustrated in Figure 2. In relation to the QUBO problem, each qubit on the QPU represents such a QUBO variable and couplers between qubits represent the costs associated with qubit pairs, mathematically described in matrix $Q$. If the problem structure can not be embedded directly to the chimera graph, auxiliary qubits may be introduced to augment the available couplings.

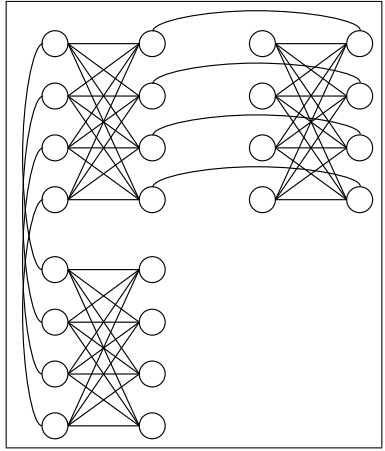

Figure 2: Excerpt of the structure of a chimera graph. The full $2048$ qubit graph extends to a $16 \times 16$ lattice of groups of $8$ qubits. Figure in reference to [2].

If – like in this paper – large data sets are used, the size of the resulting QUBO problem may exceed the limited number of available qubits on the QPU and the problem cannot be put on the chip altogether anymore. For this case, D-Wave provides a tool called QBSolv that splits the QUBO into smaller components and solves them sequentially on the D-Wave hardware [4]. A detailed view on the QBSolv algorithm is given in Section 5.1.

In this paper we used the D-Wave 2000Q model located in Vancouver, Canada, and we accessed the machine using D-Wave's cloud interface. The instance at hand has got a working graph with $2038$ qubits and $5955$ couplers out of the full graph with $2048$ qubits and $6016$ couplers.

## 3 Related Work

Over the last decades several families of heuristics have been proposed for solving the CVRP. They can be divided into construction heuristics, improvement heuristics and metaheuristics [25].

**Construction heuristics** try to generate a good solution gradually. In every step, they insert customers into partial tours or combine sub-tours considering some capacities and costs to generate a complete solution. One of the most fundamental construction heuristics is the Clarke and Wright savings algorithm [11], which first constructs a single tour for each customer, calculates the saving that can be obtained by merging those single customer tours, and iteratively combines the best sub-tours until no saving can be obtained anymore. **Improvement heuristics** try to iteratively enhance a given feasible solution, which is often generated by a construction heuristic. A common methodology is to replace or swap customers between sub-tours taking capacity constraints into account. Popular improvement methods can be found in [28] and [34]. **Metaheuristics** can be thought of as top-level strategies which guide local improvement operators to find a global solution. Groër et al. describe a library of local search heuristics for the (C)VRP [19]. In addition, Crispin and Syrichas propose a classical quantum annealing metaheuristic for vehicle scheduling [14]. To approximate quantum annealing on a classical computer, they use a stochastic variant called Path-Integral Monte Carlo (PIMC) to simulate the quantum fluctuations of a quantum system. In our work, the quantum annealing hardware is responsible for that. However, the complexity exists in mapping the CVRP to a format readable by the hardware.

---

[4]https://github.com/dwavesystems/qbsolv

One of the most important classical 2-Phase-Heuristics is the Sweep algorithm [17], where feasible clusters are formed by rotating a ray centered at the depot. After that the TSP is solved for each cluster. Fisher and Jaikumar also tried to solve the VRP with a cluster-first, route-second algorithm [16]. They formulated a Generalized Assignment Problem (GAP) instead of using a geometry based method to form the clusters. Bramel and Simchi-Levi described a 2-Phase-Heuristic where the seeds were determined by solving capacitated location problems and the remaining vertices were gradually included into their allotted route in a second stage [6].

However, there are similar investigations performed by the quantum computing community. In [37], the authors have studied the effectiveness of a quantum annealer in solving small instances within families of hard operational planning problems under various mappings to QUBO problems and embeddings. While their study did not produce results competitive with state-of-the-art classical approaches, they derive insights from the results, useful for the programming and design of future quantum annealers. In our work we investigate larger routing problem instances using a classical quantum hybrid method and state the effectiveness and efficiency. In [39], a tree-search based quantum-classical framework is presented. The authors use a quantum annealer to sample from the configuration space of a relaxed problem to obtain strong candidate solutions and then apply a classical processor that mantains a global search tree. They empirically test their algorithm and compare the variants on small problem instances from three scheduling domains. In general, one can see that many approaches have got a hybrid structure. That is, classical bottlenecks are outsourced to quantum computing devices that iteratively perform local quantum searches [39, 10, 20].

## 4 Concept of Hybrid Solution Method

There are numerous heuristics in the literature for solving the CVRP that all have an iterative approach. This makes it difficult to map them into a QUBO problem to be solved on a quantum annealer. In addition, there exists the classical 2-Phase-Heuristic that separates the CVRP into a clustering phase and a routing phase. Both phases can be seen as detached optimization problems, the Knapsack Problem (KP) with an additional minimization of distances between customers and the Travelling Salesman Problem (TSP), respectively. This division allows the mapping of the problems to one or two QUBO matrices. We have investigated three different approaches for the 2-Phase-Heuristic using a quantum annealer, see Figure 3. The insights of the preliminary exploration will be given in Subsection 4.1. The concept of the most suitable approach – called Hybrid Solution (HS) – will be presented in detail in Subsections 4.2 and 4.3.

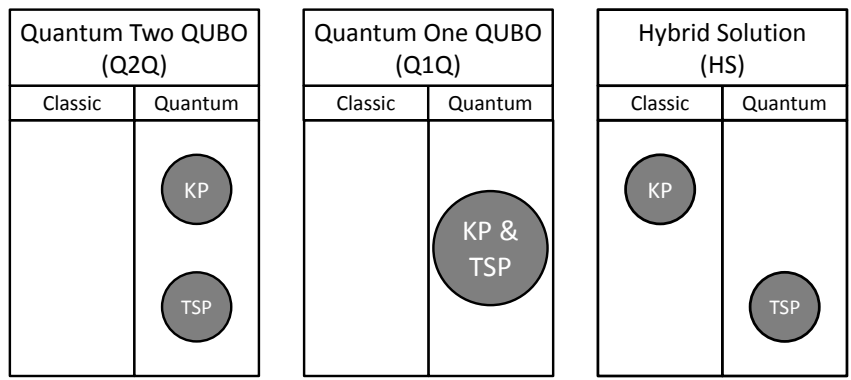

Figure 3: Assignment of the clustering phase (KP) and the routing phase (TSP) of the 2-Phase-Heuristic to a classical solution method (Classic) or quantum annealing solution method (Quantum). The algorithm on the left-hand side (Q2Q) and in the middle (Q1Q) are results of a preliminary study. The algorithm on the right-hand side (HS) is the method proposed in this paper.

### 4.1 Preliminary Exploration

The first approach based on the 2-Phase-Heuristic divides the CVRP into two separate optimization problems (Q2Q in Figure 3). The clustering phase can be considered as the Knapsack Problem with additional distance minimization between the customers to be grouped. The routing phase can be reduced to the familiar Traveling Salesman Problem. Thus, it is possible to split the CVRP down to two individual QUBO problems and execute them sequentially. The QUBO formulation for the routing phase corresponds exactly to the representation of the TSP presented in [29]. The QUBO formulation of the clustering phase, however, is an adaption of the Knapsack Problem as stated in [29]. It is composed of $H = H_A + H_B + H_C$ with

$$H_A = X \sum_{k=1}^{m} \left(1 - \sum_{n=1}^{W} y_n^k\right)^2 + A \sum_{k=1}^{m} \left(\sum_{n=1}^{W} n y_n^k - \sum_{\alpha} w_\alpha x_\alpha^k\right)^2, \tag{3}$$

$$H_B = X \sum_{\alpha} \left(1 - \sum_{k=1}^{m} x_\alpha^k\right)^2 \text{ and} \tag{4}$$

$$H_C = C \sum_{k=1}^{m} \left(\sum_{(uv) \in E} D_{uv} x_u^k x_v^k\right). \tag{5}$$

As the CVRP usually needs to fill several routes $m$ (i.e. backpacks) with customers, the original formulation of [29], which takes only one backpack into account, has been adapted accordingly with $H_A$ (Eq. 3). The first term of $H_A$ ensures that the backpack has only one capacity constraint. That is, as soon as two or more $y_n$ variables are set to 1, a penalty value $X$ (very high value) is added to the solution what declares it as bad or invalid. The second term, in turn, ensures that the sum of packed objects does not exceed the specified backpack capacity from the first term. As soon as the difference is not equal to 0, the squaring and the penalty value $A$ also classify the solution as bad. $H_B$ (Eq. 4) guarantees that every object or customer must be packed in just one backpack or route. Finally, $H_C$ (Eq. 5) is an additional optimization function that tries to improve the clustering by grouping the customers that are close to each other. To do this, the distances between the customers of a cluster are summed up. $D_{uv}$ corresponds to the Euclidean distance between customers $u$ and $v$. The solution with the shortest summed distances within the clusters is the optimum of a classical clustering. The penalty value $X$ must be greater than $A$ and $A$ greater than $C$. This ensures that in fact only one capacity limit per vehicle is set and each customer is assigned to exactly one vehicle. Experimental tests for different CVRP datasets and problem sizes yielded the following correlation of the penalty values: $X = A^2$ and $A = max(D_{uv}) *$ number of customers. $C$ corresponds to an edge weighting of the clustering, which is used to optimize the clustering.

However, in this first approach we have faced severe difficulties within the clustering phase. The problem was to find values for the edge-weighting parameter $C$ such that customers having a short distance to each other (low cost) are grouped together. Experimental results have shown that the edge-weighting parameter needs to be chosen individually for each dataset. Thus, we perceive this approach as impractical.

The second approach attempts to solve each of the CVRP subproblems simultaneously instead of sequentially (Q1Q in Figure 3). For this, the sub-problems have to be mapped to a single QUBO problem. The overall solution for this QUBO is as follows: $H = H_A + H_{B'} + H_C + H_D + H_E + H_F$ with

$$H_{B'} = X \sum_{\alpha} \left(1 - \sum_{k=1}^{m} \sum_{j=1}^{n} x_{\alpha,j}^k\right)^2, \tag{6}$$

$$H_D = X \sum_{j=1}^{n} \left(1 - \sum_{k=1}^{m} \sum_{\alpha} x_{\alpha,j}^k\right)^2, \tag{7}$$

$$H_E = E \sum_{j=1}^{N} D_{uv} x_{u,j} x_{v,j+1} \text{ and} \tag{8}$$

$$H_F = X \sum_{k=1}^{m} \left(1 - \sum_{\alpha}^{d} x_{\alpha}^{k}\right)^2. \tag{9}$$

$H_A$ corresponds to Eq. 3 of the first approach. The first term also ensures that the $m$ vehicles have a clear capacity constraint while the second term ensures that the vehicle capacity determined by the first term is not exceeded by the summed customer demand. $H_{B'}$ (Eq. 6) is similar to Eq. 4 of the first approach. However, the position in a route has additionally to be taken into account here, since in this approach both the clustering and the routing within the clusters are solved simultaneously. Thus, this term means that a customer can only be assigned exactly one route with exactly one position within this route. $H_C$ corresponds to Eq. 5 of the first approach and is responsible for optimizing the clustering. That is, we try to assign customers who have a small distance from each other to a route. $H_D$ (Eq. 7) ensures that each position $j$ can only be assigned to exactly one customer $\alpha$ and one route $k$. Finally, the shortest route within a cluster must be found. This is optimized with $H_E$ (Eq. 8). Since each position is unique across all routes, the route subdivision can be neglected here. It should also be noted that in the Q1Q approach the depot needs to be mapped to each cluster in order to properly execute the TSP in each cluster. HF causes the depots that were inserted multiple times in the dataset to be assigned to different clusters each, where $d$ corresponds to the number of depots or the number of vehicles.

However, in the second approach we observed that both optimization functions ($H_C$ and $H_E$) seemed to hinder each other. Neglecting the clustering optimization function ($H_C$) led to valid routes inside the clusters, but at the same time the clusters were very sparse. The other way round, i.e. neglecting the routing optimization function $H_E$, led to dense clusters but also to invalid routes inside the clusters. In summary, both efforts lead to invalid or unusable solutions.

The third approach (HS in Figure 3) as a candidate for a CVRP solution method combines the positive aspects of the previous mentioned approaches. To achieve this, the clustering phase (KP) is solved using a classical algorithm while the routing phase (TSP) is mapped to a QUBO problem in order to solve it on the quantum annealer. The following Subsections will go into detail.

### 4.2 Hybrid Solution – Clustering Phase

The clustering phase of the hybrid solution method we propose is inspired by Shin and Han [38]. Based on their work, we add a charateristics called *clustering core point parameter* that will be presented below. The clustering phase can be subdivided: (1) cluster generation and (2) cluster improvement.

Within the **cluster generation** the core stop of a cluster, i.e. the first customer in a cluster, is chosen. [23] propose to select the core stop either based on the maximum demand of the customers, or based on the largest distance to the depot. The motivation behind choosing the customer with the highest demand is the assumption that this one is the most critical customer in relation to the vehicles' capacity constraint. After selecting that particular customer, the vehicle can be filled with goods for customers having a smaller demand. The motivation behind choosing the customer with the largest distance to the depot is the assumption that this one is the most critical customer in relation to the routes' length constraint and that other customers may be supplied while approaching or receding that particular customer. Once the core stop $v$ of a cluster has been selected, the geometric center $CC(m_k)$ of cluster $m_k$ is calculated using

$$CC(m_k) = \sum_{i=0}^{n} \frac{v_i^x}{n}, \sum_{i=0}^{n} \frac{v_i^y}{n} \tag{10}$$

with $v_i^x$ and $v_i^y$ being the $x$ and $y$ coordinates of customer $v_i$ and $n$ being the number of customers within cluster $m_k$. Now, the customer with the smallest distance to the cluster center is selected from the set of unclustered customers and added to the cluster. After the cluster center is recalculated the steps are repeated until the demand of a customer to be added would exceed the vehicle's capacity. If this is the case, a new core stop is selected based on the previously explained criteria and the still unclustered customers are assigned to the new cluster. This procedure stops when each customer has been assigned to a cluster.

Once the clusters have been generated, the **cluster improvement** is executed to enhance the clusters. This step assigns a customer $v_i$ belonging to cluster $m_k$ to another cluster $m_j$, if that would reduce the distance to the cluster center, i.e. if the distance to $CC(m_j)$ is smaller than the distance to $CC(m_k)$. However, assigning a customer to a new cluster must not violate the capacity limitation. If the reassignment is possible and valid, $CC(m_j)$ and $CC(m_k)$ are recalculated and the improvement process begins again. The improvement step terminates if it is not possible to assign a customer to another cluster or when a certain stop criterion is reached (e.g., number of iterations).

### 4.3 Hybrid Solution – Routing Phase

After the clustering phase is completed, the goal is now to find the shortest route inside each cluster. Thus, the Travelling Salesman Problem (TSP) is executed for every generated cluster. The TSP can be reduced to the Hamiltonian Cycle Problem (HPC), which can be formulated as QUBO problem as follows [29]:

$$H_A = A \sum_{i=1}^{n} \left(1 - \sum_{j=1}^{n} x_{i,j}\right)^2 + A \sum_{j=1}^{n} \left(1 - \sum_{i=1}^{n} x_{i,j}\right)^2 + A \sum_{(ui) \notin E} \sum_{j=1}^{n} x_{u,j} x_{i,j+1} \quad (11)$$

The binary variable $x_{i,j}$ is 1 if the customer with index $i$ is located at position $j$ in the Hamiltonian Cycle. The first term (constraint) requires that each customer must occur only once in the cycle, while the second term enforces that each position in the cycle must be assigned to exactly one customer. This defines the order of the customers within the tour. The squared differences of these terms ensure that exactly one customer has a unique position in the tour. Otherwise, a high penalty value $A$ would be added to the solution energy, which states the solution itself as suboptimal or rather invalid. Although different penalties are possible for the two terms, we choose the same value because we consider both constraints equally important. The third term ensures that the order of customers found is possible. That is, if $x_{u,j}$ and $x_{i,j+1}$ are both 1 and $(ui) \notin E$ with $E$ being the set of edges between the nodes representing the customers, then the penalty value $A$ should also state the solution invalid [29]. In this work we have evaluated our algorithm using CVRP/TSP datasets with fully meshed vertices, i.e. customers are connected with undirected edges. That is why the third term can be neglected.

In order to find the Hamiltonian Cycle with the shortest length, the following minimization function is needed:

$$H_B = B \sum_{(ui) \in E} D_{ui} \sum_{j=1}^{n} x_{u,j} x_{i,j+1} \quad (12)$$

Here $D_{ui}$ is the euclidean distance between the customers $u$ and $i$. The minimization function sums all costs of the edges between successive customers. The total solution for the TSP QUBO problem is then composed of:

$$H = H_A + H_B \quad (13)$$

The penalty value $B$ must be chosen sufficiently small so that it does not violate the constraint $H_A$. A possible choice would be $0 < B \cdot max(D_{ui}) < A$ [29]. With $B = 1$, $A$ has to be chosen larger than the greatest distance between two customers. In our experiments $B$ was set to 1 and $A$ was set to $n \cdot max(D_{ui})$ with $n$ being the number of customers.

By multiplying the QUBO formulas, one obtains the coefficients for the QUBO problem matrix which can be written as a lower (or upper) triangular matrix to be mapped to the quantum annealing processor. Figure 4 shows an excerpt of an exemplary QUBO problem in which only the coefficients are entered for simplification. As soon as several coefficients are noted on one cell of the matrix they must be added. In addition, every coefficient is multiplied with the penalty value $A$ and the distance is multiplied with penalty value $B$.

|    | A1 | A2 | A3 | A4 | B1 | B2 | B3 | B4 | C1 | C2 | C3 | C4 |
|----|----|----|----|----|----|----|----|----|----|----|----|----|
| A1 | -1/-1 | 2 | 2 | 2 | 2 | $D_{ab}$ | 0 | $D_{ba}$ | 2 | $D_{ac}$ | 0 | $D_{ca}$ |
| A2 |    | -1/-1 | 2 | 2 | $D_{ba}$ | 2 | $D_{ab}$ | 0 | $D_{ca}$ | 2 | $D_{ac}$ | 0 |
| A3 |    |    | -1/-1 | 2 | 0 | $D_{ba}$ | 2 | $D_{ab}$ | 0 | $D_{ca}$ | 2 | $D_{ac}$ |
| A4 |    |    |    | -1/-1 | $D_{ab}$ | 0 | $D_{ba}$ | 2 | $D_{ac}$ | 0 | $D_{ca}$ | 2 |
| B1 |    |    |    |    | -1/-1 | 2 | 2 | 2 | 2 | $D_{bc}$ | 0 | $D_{cb}$ |
| B2 |    |    |    |    |    | -1/-1 | 2 | 2 | $D_{cb}$ | 2 | $D_{bc}$ | 0 |
| B3 |    |    |    |    |    |    | -1/-1 | 2 | 0 | $D_{cb}$ | 2 | $D_{bc}$ |
| B4 |    |    |    |    |    |    |    | -1/-1 | $D_{bc}$ | 0 | $D_{cb}$ | 2 |
| C1 |    |    |    |    |    |    |    |    | -1/-1 | 2 | 2 | 2 |
| C2 |    |    |    |    |    |    |    |    |    | -1/-1 | 2 | 2 |
| C3 |    |    |    |    |    |    |    |    |    |    | -1/-1 | 2 |
| C4 |    |    |    |    |    |    |    |    |    |    |    | -1/-1 |

Figure 4: Excerpt of a visualized TSP QUBO problem matrix: A1 corresponds to customer A on position 1 in the TSP cycle. The colored entries correspond to the coefficients and distances from Equation 11 and 12 , respectively.

## 5 Evaluation

In this section we present the results of the hybrid solving method with regards to solution quality and computational results. First the preliminaries for the test setup are given, then the test results are shown.

### 5.1 Preliminaries

As already mentioned, the D-Wave quantum annealer can be regarded as a discrete optimization machine that accepts problems in QUBO formulation. The QUBO problem matrix increases with the problem size, i.e. with the number of problem variables used. For the TSP $n^2$ logical variables, with $n$ being the number of customers, have to be used to describe it as a QUBO problem (see Section 4.3). These variables have to be mapped to the qubits and the logical links between them to the couplers of the physical QPU. Because of the almost fully meshed dependencies between the logical variables it is not possible that the logical problem structure matches the physical one. For such issues D-Wave provides a minor embedding technique to find a valid embedding to the hardware, as described in [8]. We have used this technique [5] in combination with D-Wave's QBSolv tool to fit our large QUBO problems to the physical hardware.

QBSolv splits the QUBO into smaller components (subQUBOs) of a predefined subproblem size [6], which are then solved independently of each other. This process is executed iteratively as long as there is an improvement and it can be defined using the QBSolv parameter *num_repeats*. This parameter determines the number of times to repeat the splitting of the QUBO problem matrix after finding a better sample. With doing so, the QUBO matrix is split into different components using a classical tabu search heuristic in each iteration. QBSolv can be used in a completely classical way to solve the subQUBOs or as a quantum-classic hybrid method by solving the single subQUBOs on the quantum annealer.

Besides embedding and splitting the QUBO into subQUBOs, QBSolv also takes care of the unembedding and the merging of the subproblems' solutions. We use the default configuration of D-Wave's QBSolv [7] including the *auto_scale* function that automatically scales the values of the QUBO matrix to the allowed range of values for the biases and strenghts of qubits and couplers. The single-shot annealing time is set to the default value of $20\mu s$. For more details about QBSolv, see [32].

---

[5] https://github.com/dwavesystems/qbsolv/blob/master/examples/useFixedEmbeddingComposite.py
[6] Due to experimental tests we have chosen a subQUBO size of 20 logical variables.
[7] To the time of writing QBSolv's current version 0.2.8.

There exist many different benchmark datasets for the CVRP and the TSP, which can be downloaded from [40], [36], [12]. In addition, the *Best Known Solution* (BKS) of each dataset is noted. It gives information about the best solution, i.e., the shortest euclidean distance found by any solution method. In order to test and compare the proposed hybrid solution method with regard to the solution quality, various test datasets of Christofides and Eilon [40] have been selected. Details about the CVRP datasets can be extracted from the name with the format *E-nX-kY*. For example, *E-n22-k4* stands for a certain dataset $E$, $n22$ for the number of customers including the depot and $k4$ for the minimal number of vehicles needed to solve the problem. The TSP datasets have the name format *CityX*, which just indicates the number of customers which have to be visited in the TSP tour. As already mentioned, the customer and depot coordinates relate to a 2D euclidean space.

## 5.2 Results

In this subsection the results of our hybrid method are presented. First, we exclusively analyze the TSP which is executed on the quantum annealer to see how the different-sized problem instances are handled.

### 5.2.1 TSP – Solution Quality

Table 1 shows the results for different-sized TSP datasets. The problem instance and its size, also included in the name, can be read from column one and two. In addition, the BKS, the best solution of 100 runs, and the average deviation of all 100 runs from the BKS, is noted in column three to five, respectively. One can see that for smaller sized problem instances (1)(2) the BKS has been found and the average deviation is generally low (0.00% to 0.31%). With increasing problem size (3)(4)(5) the BKS is not found and the average deviation increases continuously (2.70% to 25.91%). Therefore it can be concluded that the TSP can be solved comparatively well for smaller sized problem instances on the quantum annealer, while with regard to larger sized instances only a good approximation is found.

| Problem | Size | Best Known Solution (BKS) | Best solution found of 100 runs | Avg. deviation of 100 runs from BKS |
|---|---|---|---|---|
| (1) Burma14 | 14 | 3323 | 3323 | 0.00% |
| (2) Ulysses16 | 16 | 6859 | 6859 | 0.31% |
| (3) Ulysses22 | 22 | 7013 | 7019 | 2.70% |
| (4) WesternSahara29 | 29 | 27603 | 28293 | 8.16% |
| (5) Djibouti38 | 38 | 6656 | 7396 | 25.91% |

Table 1: Results for various TSP datasets. Parameter num_repeats was set to 250.

In Figure 5 the deviation of each found solution from the BKS is visualized with a boxplot diagram. In this figure the variance of each test result in relation to the deviation is shown in more detail. It also seems that with increasing problem size the variance of the results expands. Whereas for datasets (1)(2)(3) the variance is comparatively low (0.00%, 0.00%-1.56%, and 0.09%-5.29%), larger datasets (4)(5) show more fluctuations (2.50%-13.77% and 11.12%-36.01%).

Experimental tests showed that the solution quality depends on the QBSolv parameter *num_repeats*. In Figure 6 the deviation from the BKS for the already known datasets with different settings for the *num_repeats* parameter are shown. Dataset *Burma14* has been neglected since even *num_repeats* set to 50 finds the BKS in every run. Each setting has been executed 10 times. One can see that with increasing the *num_repeats* parameter there is a tendency that the solution quality improves, i.e., the deviation from the BKS decreases.

### 5.2.2 CVRP – Solution Quality

Now the results of the hybrid method including its classical clustering phase are presented. One can optimize the clustering of the customers by choosing the core stop of a cluster (*max_distance* or *max_request*). Depending on the selected parameter, either the customer with the largest distance to the depot or the customer with the highest demand is set as the seed of a cluster. In Table 2 the best solution found is noted, i.e. the sum of all vehicle routes and the deviation from the BKS for both options of the core stop parameter.

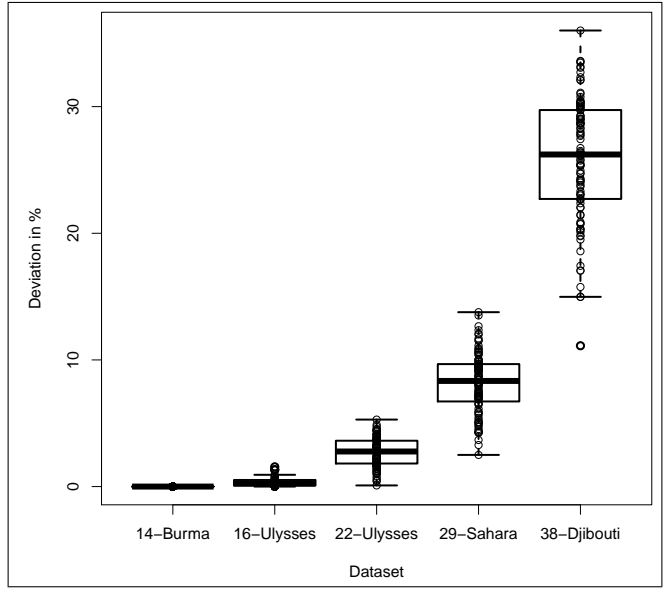

Figure 5: Deviation of the results from the BKS for each dataset. Every dataset was run 100 times and num_repeats was set to 250. The dots represent the measured deviations. The box corresponds to the area in which the middle 50% of the data reside in with the continuous line being the median. The whiskers extend to the most extreme data point which is no more than 1.5 times the interquartile range from the box.

| | | Hybrid Solution Method | | | |
| --- | --- | --- | --- | --- | --- |
| Problem | BKS | Shortest distance (max_distance) | Shortest distance (max_request) | Avg. deviation from BKS (max_distance) | Avg. deviation from BKS (max_request) |
| (1) E-n22-k4 | 375 | 385 | 407 | 2.66% | 8.53% |
| (2) E-n33-k4 | 835 | 965 | 852 | 15.07% | 2.05% |
| (3) E-n51-k5 | 521 | 618 | 557 | 18.75% | 6.91% |
| (4) E-n76-k7 | 682 | 748 | 814 | 9.70% | 19.46% |
| (5) E-n101-k8 | 815 | 892 | 898 | 11.16% | 11.39% |

Table 2: Distances and deviations from the BKS for certain E datasets of Christofides and Eilon. Every dataset was executed 10 times.

The results do not allow a concrete statement about the choice of the core stop parameter of the hybrid method. One can see, that it is independent of the problem size. In fact the core point has to be selected individually for each dataset to obtain a good clustering and a short route length as a consequence. Especially for the first three datasets (1)(2)(3) the hybrid method finds good approximations to the BKS with regard to the solution quality (2.66%-6.91% deviation).

In Table 3 the solution quality of the hybrid method is compared to other fundamental construction and 2-phase-heuristics. The solution quality has been compared based on seven CMT datasets of Christofides, Mingozzi and Toth [40]. Datasets CMT6-CMT10 have been neglected in the evaluation as they are identical to CMT1-CMT5, however, they regard an additional time window, which in turn is ignored in the classic CVRP. The solutions for the selected datasets found by the other heuristics were extracted from [16]. For every solution method the best solution found (Distance) and the deviation (Dev.) from the BKS is depicted. In addition, the problem size and the BKS for each dataset is noted in column two and three.

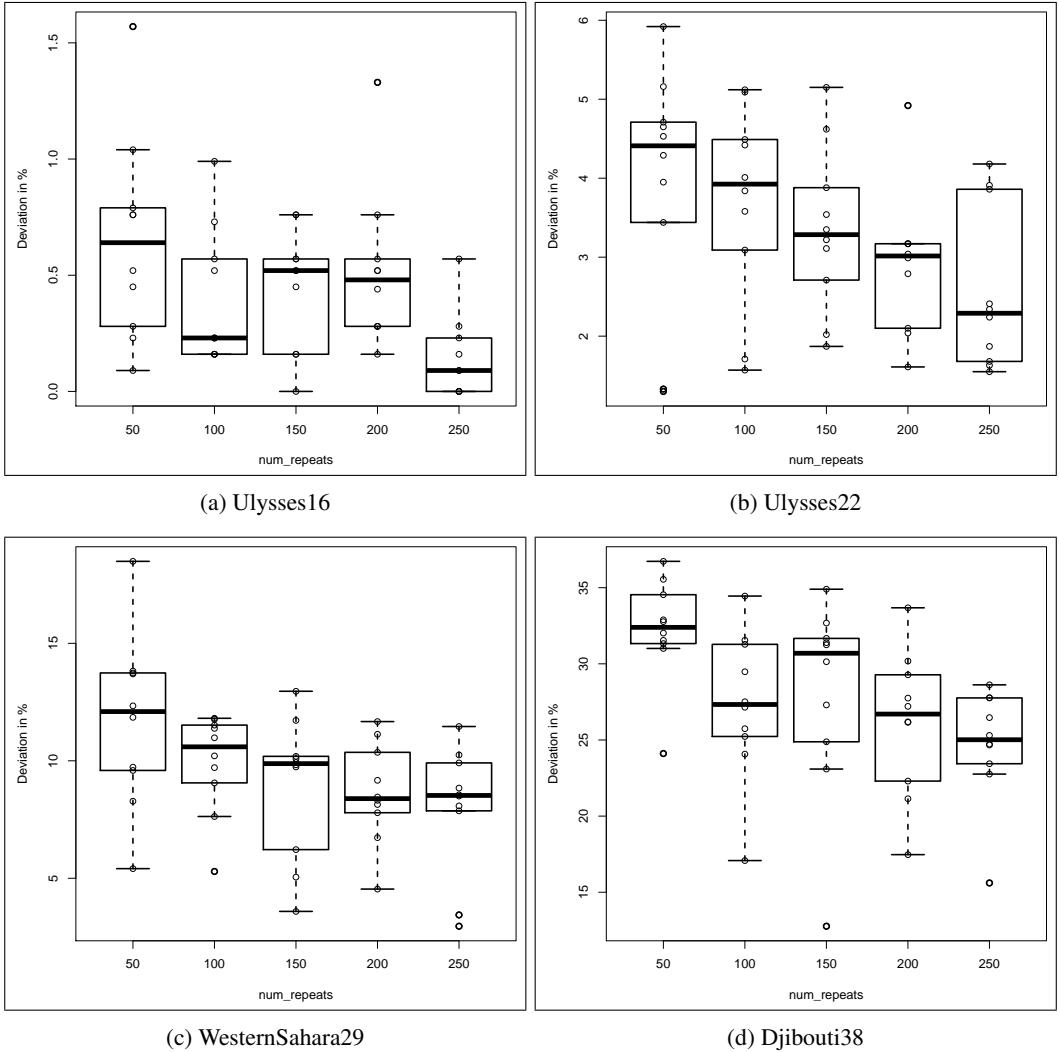

Figure 6: Different settings for the *num_repeats* parameter of QBSolv for various datasets. Meaning of dots, box and whiskers as described with Figure 5.

For problem instances CMT1 and CMT5, the hybrid method was able to keep up or even surpass the Saving-Heuristic of Clarke and Wright. With regard to datasets CMT2, CMT3 and CMT4 major deviations from the BKS are recorded. Here the hybrid method can keep up with or even dominate the Saving-Heuristic. However, with regard to the other heuristics it was not competitive. The last two problem instances are structured problems in which the customers are organized in clusters by the authors of the datasets. According to Fisher and Jaikumar, these datasets resemble real problems rather than problem instances CMT1-CMT5 [16]. Therefore, with these instances one can recognize that the clustering algorithm of the hybrid method works comparatively well, since each of the other four heuristics was surpassed with regard to the solution quality. However, in relation to the BKS it should be mentioned that both, the hybrid method and the other heuristics, never found the BKS (exception: problem instance CMT1 - Fisher and Jaikumar). This is basically a well-known problem of clustering customers in the CVRP, since finding the BKS or the global optimum is related to the geographical structure of the customers of a problem instance.

### 5.2.3 CVRP – Computational Results

The computation time for the executed test instances must be considered differentiated. As already mentioned, QBSolv can be used as a pure classical solver (called *Local*) as well as a hybrid solver (called *Remote*) for large QUBO problems. With the classic version, the subQUBOs are solved locally, while with the hybrid version the subQUBOs are solved sequentially on the D-Wave hardware. Doing this, the QUBO is split locally and the subQUBOs are sent to the hardware via a remote connection and the individual jobs are placed in a queue. As a result of this process, additional latency and possibly waiting times may occur.

| | | | Clarke-Wright | | Fisher-Jaikumar | | Christofides et al. 2 Phase | | Sweep | | Hybrid Method | |
|---|---|---|---|---|---|---|---|---|---|---|---|---|
| Problem | Size | BKS | Distance | Dev. | Distance | Dev. | Distance | Dev. | Distance | Dev. | Distance | Dev. |
| CMT1 | 50 | 524.61 | 585 | 11.5% | 524 | 0.12% | 550 | 4.84% | 532 | 1.41% | 556[b] | 5.98% |
| CMT2 | 75 | 835.26 | 900 | 7.75% | 857 | 2.60% | 883 | 5.72% | 874 | 4.64% | 926[a] | 10.86% |
| CMT3 | 100 | 826.14 | 886 | 7.25% | 833 | 0.83% | 851 | 3.01% | 851 | 3.01% | 905[a] | 9.55% |
| CMT4 | 150 | 1028.42 | 1204 | 17.07% | - | - | 1093 | 6.28% | 1079 | 4.92% | 1148[a] | 11.63% |
| CMT5 | 199 | 1291.29 | 1540 | 19.26% | 1420 | 9.97% | 1418 | 9.81% | 1389 | 7.57% | 1429[a] | 10.66% |
| CMT11 | 120 | 1042.12 | - | - | - | - | - | - | - | - | 1084[b] | 4.02% |
| CMT12 | 100 | 819.56 | 877 | 7.01% | 848 | 3.47% | 876 | 6.89% | 949 | 15.79% | 828[a] | 1.03% |

Table 3: Results for selected CMT datasets: Solution found with (a) max_distance or (b) max_request

To demonstrate the difference in relation to the computational time of the routing phase, the CMT1 dataset has been used. For each of the 5 contained clusters, the routing phase has to find the shortest tour visiting all clusters' customers. This has been done locally as well as remotely on the quantum annealer using QBSolv. A distance of 557 has been found in the classical way and the hybrid version found a distance of 556. The corresponding computational results are consolidated in Table 4. The numbers of the table are based on the listings given in the appendix: In Listing 1 and Listing 2 the measured CPU times for the locally executed QBSolv are shown, while Listing 3 and Listing 4 show the same for the remotely executed QBSolv. The dataset has been run on a Dell 2.8 GHz i7 with 16 GB RAM Notebook. For measuring the CPU processing time the Python module cProfile has been used.

| | Local | | | Remote | | |
|---|---|---|---|---|---|---|
| | CPU | QPU | Total | CPU | QPU | Total |
| Cluster 1 | 0.016 | - | 0.016 | 3.281 | 0.031 | 3.312 |
| Cluster 2 | 0.046 | - | 0.046 | 2.016 | 0.016 | 2.032 |
| Cluster 3 | 0.016 | - | 0.016 | 1.235 | 0.016 | 1.251 |
| Cluster 4 | 0.031 | - | 0.031 | 3.535 | 0.031 | 3.566 |
| Cluster 5 | 0.125 | - | 0.125 | 1.184 | 0.016 | 1.200 |
| Sum QBSolv | 0.234 | - | 0.234 | 11.251 | 0.110 | 11.361 |
| Main Procedure | 1.240 | - | 1.240 | 4.431 | - | 4.431 |
| Total Runtime | | | 1.474 | | | 15.792 |

Table 4: Computational results for the CMT1 dataset. Various time components (CPU, QPU and Total Time) are given for locally and remotely used QBSolv. Cluster 1-5 correspond to the 5 clusters of the CMT1 dataset found by the clustering phase of our Hybrid Solution.

The total runtime of the locally executed algorithm consists of two parts, the main procedure (clustering phase, QUBO construction, I/O) and the actual routing (i.e., the QBSolv runtime). In Table 4 can be seen, that the total runtime of the classically executed hybrid solution algorithm took 1.474 seconds to complete consisting of a QBSolv runtime of 0.234 seconds and a main procedure of 1.24 seconds. In contrast to that, the remote version of our hybrid solution method needed 15.792 seconds. This is due to the fact, that the algorithm additionally encounters times for embedding the problem onto the Chimera graph, latency of the Internet connection and queueing at the quantum hardware. The method listing of cProfile in Listings 2 and 4 clearly show that the hybrid version of QBSolv mainly stays in the method *method 'acquire' of 'thread.lock' objects*, which is not listed in the locally QBSolv run. Therefore we assume that here the main QBSolv thread waits for the child threads to find a valid embedding for the respective subQUBOs to the D-Wave hardware. This process is not needed using QBSolv locally. However, the actual annealing time for solving the QUBO problem remotely is in the range of 0.016 to 0.031 seconds. Please note, that the QBSolv method splits the 5 problem QUBOs into 4, 2, 2, 4 and 2 subQUBOs (*Number of Partitioned Calls* and solves them one after the other on the quantum annealer. Thus, the listed numbers in Table 4 are based on the QPU access times per subQUBO in microseconds pictured in Listing 5. These add up to 0.11 seconds. The classic QBSolv version requires 0.234 seconds to solve the 5 QUBOs.

In summary, it can be stated that the real computation time to solve a QUBO problem on D-Wave's quantum annealer is comparatively shorter than with the classic QBSolv version (0.234 seconds locally versus 0.11 seconds remotely). However, methods like finding a valid embedding of the QUBO problem to the hardware, which is not needed for using QBSolv locally, generates overhead. For this reason, the classic version of QBSolv is currently more advantageous in practice regarding the total calculation time.

## 6 Conclusion

To the best of our knowledge, this work presents the first study that solves the capacitated vehicle routing problem (CVRP) using quantum annealing hardware. The most important step was to find an intuitive way to map this optimization problem to a QUBO problem. Doing this, the classical 2-phase-heuristic has been used, which enables to divide the complex problem into a clustering phase as well as a routing phase and solves them sequentially or simultaneously (see approaches Q2Q and Q1Q in Figure 3). Due to various problems within the clustering phase, a hybrid method proved to be the best. We showed that our hybrid method was able to compete with other classical construction and 2-phase-heuristics and in some cases even surpass them with regard to solution quality. However, it should be mentioned that there are other solution methods like metaheuristics, which provide a better solution with regard to the used benchmark datasets. Especially when using datasets whose BKS contains overlapping routes, the hybrid method – and in general heuristics with clustering methods that work distance-based – has got difficulties in finding the BKS. However, the hybrid method usually provided a good approximation.

The results of the present study can also be considered detached from the CVRP domain. One part of the hybrid solution method has been formalized as the TSP. As a decision problem, the TSP is known to be NP-complete. However, practically relevant tasks using the TSP as a building block require solving the optimization variant of the problem [5]. TSP is known to be APX-complete [13], which essentially means it is hard to approximate – any efficient polynomial time algorithm can only find answers that differ from the optimal solution by a constant multiplicative factor. The question now is whether a quantum annealing device can offer advantages over classical algorithms. On the one hand, the advantage can relate to the solution quality, as already described above. On the other hand, the advantage can be regarding the time to solution.

Thus, the computational time of the hybrid solution method must be considered differentiated. Due to the currently limited number of available qubits on the D-Wave QPU, the tool QBSolv must be used for large QUBO problems which are not directly embeddable on the D-Wave chip. This tool makes it possible to split the QUBO into smaller subQUBOs and place them one after the other on the quantum annealer. However, this hybrid solution option involves certain latency and waiting times which lacks the hoped acceleration of the computational time compared to the classical option. Thus, with an increasing size of the hardware the necessity of using QBSolv is decreasing. This results in a drastically lowered classical overhead time while the real solution time for an embeddable QUBO problem on the D-Wave quantum annealer is expected to stay in the range of microseconds.

In summary, the hybrid solution method presented in this study has not brought clear benefit in solution quality or computational time. Nonetheless, we have presented an approach on how to split complex combined problems and solve them in a hybrid way using a quantum annealer. This in turn can serve as a basis for further optimization problems. Working out a clear advantage in terms of time or quality thus remains future work, as we wait for larger hardware. Part of the future work will then be to investigate the effective scaling: how does the size of the hardware affects the efficiency of the problem mapping, the necessity of using additional tools like QBSolv, but also the duration of the annealing process itself.

At the 2018 D-Wave Qubits Europe users conference D-Wave provided an outlook about the future hardware directions of quantum annealing. They stated that the connectivity and the number of qubits on D-Wave machines will immensely rise over the next years [8]. These news give hope that in the future D-Wave's quantum annealers are more suitable for these kind of optimization problems and a shorter total computation time can be achieved.

---

[8] https://www.dwavesys.com/sites/default/files/mwj_dwave_qubits2018.pdf

## Acknowledgement

Research was funded by Volkswagen Group, department Group IT.

# Appendix

```
1   number of clusters: 5
2
3   001000000000010000000100000000000010000000000001000000100000000000010000001000000100001000000000
4   -4922.00000 Energy of solution
5   4 Number of Partitioned calls, 1 output sample
6   0.01600 seconds of classic cpu time
7
8   000000001000000000100000000000000010010000000000000000001001000000001000000000000001000000000000100000
9   -7646.00000 Energy of solution
10  4 Number of Partitioned calls, 1 output sample
11  0.04600 seconds of classic cpu time
12
13  000100000000000010000000000001000000000100000000000001000000100000000000000010000000000000010000
14  -12552.00000 Energy of solution
15  2 Number of Partitioned calls, 1 output sample
16  0.01600 seconds of classic cpu time
17
18  000000100000000001000000010000000000001000100000001000000000001000000100000000000001000000000000001
19  -6903.00000 Energy of solution
20  8 Number of Partitioned calls, 1 output sample
21  0.03100 seconds of classic cpu time
22
23  000100000000010000000010000000000000010000000000010000000000000000100000000010000000001000000
24  -16253.00000 Energy of solution
25  32 Number of Partitioned calls, 1 output sample
26  0.12500 seconds of classic cpu time
27
28  Total distance 557
```

Listing 1: Output of locally used QBSolv.

```
1   738180 function calls (661690 primitive calls) in 1.474 seconds
2
3   Ordered by: internal time
4
5   ncalls  tottime  percall  cumtime  percall  filename:lineno(function)
6   5       0.262    0.052    0.262    0.052    {dwave_qbsolv.qbsolv_binding.run_qbsolv}
7   51      0.066    0.001    1.258    0.025    __init__.py:1(<module>)
8   364     0.058    0.000    0.058    0.000    {imp.find_module}
9   37752   0.056    0.000    0.056    0.000    binary_quadratic_model.py:484(add_interaction)
10  5       0.043    0.009    0.044    0.009    TSPPreparer.py:56(generate_adjacency_matrix)
11  78703   0.040    0.000    0.060    0.000    records.py:438(__getattribute__)
12  5       0.033    0.007    0.033    0.007    TSPPreparer.py:25(generate_edge_list_tsp)
13  5       0.026    0.005    0.028    0.006    TSPPreparer.py:98(generate_edge_matrix)
14  705     0.025    0.000    0.025    0.000    {compile}
15  5       0.023    0.005    0.078    0.016    binary_quadratic_model.py:595(add_interactions_from)
16  4       0.020    0.005    0.202    0.050    __init__.py:4(<module>)
17  11195   0.020    0.000    0.093    0.000    response.py:675(__getitem__)
18  5       0.016    0.003    0.018    0.004    TSPSolver.py:11(generate_tsp_qubo)
19  ...
```

Listing 2: Output of cProfile for locally used QBSolv.

```
1   number of clusters: 5
2
3   00000000011000000000000000010010000000000010000001000000000010000001000000100000000000100
4   -4922.00000 Energy of solution
5   4 Number of Partitioned calls, 1 output sample
6   3.31200 seconds of classic cpu time
7
8   001000000000010000000100000000000000010100000000000010000000000000100000010000000100000000
9   -7646.00000 Energy of solution
10  2 Number of Partitioned calls, 1 output sample
11  2.03200 seconds of classic cpu time
12
13  00000001000000000010000000000000001000000100000000010000000000000001000010000000000000100100
14  -12551.00000 Energy of solution
15  2 Number of Partitioned calls, 1 output sample
16  1.25100 seconds of classic cpu time
17
18  000100000000100000000000100000010000000000000100000000010000000010000000001000001000100000000
19  -6903.00000 Energy of solution
20  4 Number of Partitioned calls, 1 output sample
21  3.56600 seconds of classic cpu time
22
23  00010000000010000000001000000000000001000000000000010000000001000000001000000001000000
24  -16246.00000 Energy of solution
25  2 Number of Partitioned calls, 1 output sample
26  1.20000 seconds of classic cpu time
27
28  Total distance 566
```

Listing 3: Output of remotely used QBSolv.

```
1   882498 function calls (821251 primitive calls) in 15.792 seconds
2
3   Ordered by: internal time
4
5   ncalls  tottime  percall  cumtime  percall  filename:lineno(function)
6   241     11.205   0.046    11.205   0.046    {method 'acquire' of 'thread.lock' objects}
7   5023    1.028    0.000    1.028    0.000    {built-in method __new__ of type object at 0x53A271E0}
8   55      0.926    0.017    0.926    0.017    {method 'read' of '_ssl._SSLSocket' objects}
9   1       0.618    0.618    0.618    0.618    {minorminer.find_embedding}
10  1       0.368    0.368    0.368    0.368    {method 'do_handshake' of '_ssl._SSLSocket' objects}
11  1       0.181    0.181    0.181    0.181    {method 'connect' of '_socket.socket' objects}
12  5       0.077    0.015    11.369   2.274    {dwave_qbsolv.qbsolv_binding.run_qbsolv}
13  51      0.068    0.001    1.287    0.025    __init__.py:1(<module>)
14  41476   0.065    0.000    0.065    0.000    binary_quadratic_model.py:484(add_interaction)
15  5       0.044    0.009    0.045    0.009    TSPPreparer.py:56(generate_adjacency_matrix)
16  5       0.035    0.007    0.035    0.007    TSPPreparer.py:25(generate_edge_list_tsp)
17  61672   0.033    0.000    0.049    0.000    records.py:438(__getattribute__)
18  3       0.031    0.010    0.032    0.011    decoder.py:370(raw_decode)
19  701     0.026    0.000    0.091    0.000    binary_quadratic_model.py:595(add_interactions_from)
20  5       0.026    0.005    0.028    0.006    TSPPreparer.py:98(generate_edge_matrix)
21  705     0.025    0.000    0.025    0.000    {compile}
22  5       0.017    0.003    0.019    0.004    TSPSolver.py:11(generate_tsp_qubo)
23  ...
```

Listing 4: Output of cProfile for remotely used QBSolv.

```
1   qpu_access_time 7783        qpu_access_time 7789
2   qpu_access_time 7787        qpu_access_time 7783
3   qpu_access_time 7808        qpu_access_time 7807
4   qpu_access_time 7814        qpu_access_time 7778
5   qpu_access_time 7768        qpu_access_time 7795
6   qpu_access_time 7802        qpu_access_time 7788
7   qpu_access_time 7807        qpu_access_time 7782
```

Listing 5: qpu_access_time of using QBSolv remotely. The qpu_access_time includes the whole time range from programming to post processing overhead time.